\documentclass[aps,pra,twocolumn,superscriptaddress,groupedaddress]{revtex4}  % for double-spaced preprint
\usepackage{graphicx}  % needed for figures
\usepackage{dcolumn}   % needed for some tables
\usepackage{bm}        % for math
\usepackage{amssymb}   % for math
\usepackage{amsmath}
\usepackage{xcolor}
\usepackage{pdfpages}

\begin{document}

\title{Nonlinear refractive index of warm rubidium vapor}

\author{L. Kardum}
\affiliation{Centre for Advanced Laser Techniques, Institute of Physics, Bijeni\v{c}ka cesta 46, 10000 Zagreb, Croatia}
\author{G. Premec}
\affiliation{Centre for Advanced Laser Techniques, Institute of Physics, Bijeni\v{c}ka cesta 46, 10000 Zagreb, Croatia}
\author{N. \v{S}anti\'{c}}
\affiliation{Centre for Advanced Laser Techniques, Institute of Physics, Bijeni\v{c}ka cesta 46, 10000 Zagreb, Croatia}
\author{D.~Aumiler}
\affiliation{Centre for Advanced Laser Techniques, Institute of Physics, Bijeni\v{c}ka cesta 46, 10000 Zagreb, Croatia}
\email{aumiler@ifs.hr}

\date{\today}

\begin{abstract}
The potential to precisely control both the linear and nonlinear index of refraction through optical manipulation of the atomic states has recently pushed warm alkali vapors to the forefront of research in the field of quantum sensors, quantum memories, and quantum fluids of light. 
Rubidium (Rb) vapor in centimeter-scale glass cells or millimeter-scale MEMS cells has proven to be a very promising platform for these applications, yet only a handful of research works have been dedicated to the investigation of the (non)linear refractive index of Rb vapor. 
We present results of theoretical calculations of the (non)linear refractive index of warm Rb vapor, based on the optical Bloch equations for 6-level Rb atoms interacting with a probe laser. 
They are compared to the experimental results obtained using an interferometric technique, showing excellent quantitative agreement. 
A Kerr nonlinear refractive index $n_2$ of up to $10^{-4}$ cm$^2$/W is obtained. 
Python scripts for all theoretical calculations presented in this work are provided, including the refractive index calculation, that can readily be used in practical implementations for simulating the (non)linear refractive index of Rb vapor including the effects of Doppler broadening, transit time broadening, pressure broadening, saturation, optical pumping, and spin-exchange collisions.  
\end{abstract}

\maketitle

\section{Introduction}
Warm atomic vapors have recently seen a striking revival triggered by their application in chip-scale atomic devices such as atomic clocks and magnetometers \cite{kitching2018}, and in the research of squeezed light generation \cite{ries2003}, quantum memories \cite{treutlein2023}, and quantum fluids of light \cite{santic2018}. 
In all these applications warm atomic vapors are a medium of choice since they allow for precise control of the linear and nonlinear index of refraction through optical manipulation of the atomic states \cite{glorieux2023}.
Nonlinear effects in atomic vapors emerge at very low powers, already on the milliwatt power scale (see e.g. \cite{vulic2026}).
This enables the observation of nonlinear optical phenomena using low-intensity continuous wave lasers (of the order of milliwatts per square centimeter), but simultaneously means that saturation effects must routinely be considered unless very low laser intensities are used.

The simplest nonlinear process is the Kerr effect, whereby the index of refraction in a medium depends linearly on the intensity of the light passing through it.
It is however relatively easy to saturate the Kerr effect in atomic vapors close to atomic resonances, which can not be described using lowest-order perturbation theory, so the saturated Kerr coefficient has been invoked \cite{mccormick2004}:
\begin{equation}
	n(I) = n_0 + \frac{n_2 I}{1+I/I_s},
	\label{Eq:n2}
\end{equation}
where $n_2$ is the nonlinear Kerr coefficient, and $I_s$ is the saturation intensity. 

Surprisingly, only a handful of experiments regarding the measurement of the nonlinear refractive index of rubidium vapor can be found in the literature. 
In their seminal works McCormick \textit{et. al.} directly measured $n_2$ in atomic rubidium vapor using the z-scan method and discussed saturation effects \cite{mccormick2003,mccormick2004}. 
Recently, nonlinear refractive index measurements of rubidium vapor have been extended to spatial mapping onto the light beam profile \cite{wu2022}, and using the z-scan method with an optical frequency comb \cite{wang2020}. 
In the latter, a frequency comb of 0.1 nm spectral bandwidth was used resulting in the nonlinear response of the atomic medium that includes contributions of hundreds of comb modes, making the comparison to continuous wave measurements not straightforward. 
Finally, theoretical calculations of the nonlinear refractive index of rubidium vapor were recently presented based on analytical results for a two-level system \cite{levine2023}, and numerical results including all magnetic sublevels in order to include the effects of incomplete optical pumping \cite{levine2025}. 
These theoretical works have managed to reproduce the measured values only semiquantitatively, and highlighted the need for more detailed theoretical modeling as well as the improved experimental data. 

In this work we present results of a theoretical and experimental investigation of the nonlinear refractive index of warm rubidium (Rb) atomic vapor. 
We use the density matrix formalism and optical Bloch equations to calculate the optical response of multi-level Rb atoms (including hyperfine structure) to laser light. 
In addition, we provide Python scripts for all theoretical calculations presented in this work, including the refractive index calculation, that can readily be used in practical implementations for simulating the linear and nonlinear refractive index of Rb vapor including the effects of Doppler broadening, transit time broadening, pressure broadening, saturation, optical pumping, and spin-exchange collisions. 
In the experimental part, we employ an interferometric technique to measure the refractive index of $^{87}$Rb vapor as a function of frequency and laser power. 
Finally, we show excellent quantitative agreement of the theoretical model with experimental results. 

The structure of the paper is as follows. 
First we introduce the theoretical model for calculating the refractive index of rubidium atomic vapor, based on the optical Bloch equations for 6-level Rb atoms. 
The calculated refractive index is then employed to simulate the propagation of a laser beam through the vapor using the paraxial nonlinear wave equation, and to calculate the spatially varying nonlinear phase accumulated by the laser beam after passing through the vapor. 
The phase calculated in this way is compared to the experimental results obtained by using an interferometric technique that measures the local phase difference between a reference laser beam and a beam that has interacted with the Rb vapor.
Finally, we analyze the saturation of the nonlinear refractive index and provide numerical values for the nonlinear Kerr coefficient ($n_2$) and the saturation intensity ($I_s$) that are relevant for the D2 resonance line in warm $^{87}$Rb vapor.

%%%%%%%%%%%%%%%%%%%%%%%%%%%%%%%%%%%%%%%%%%%%%%%%%%%%%%%%%%%%%%%%%%%%%%%%%%%%%%%%%%%%%%%%%%%%%%%%%%%%%%%%%%%%%%%%%%%%%%%%%%%%%%%%%%%%%%%%%
\section{Theoretical model for calculating the nonlinear refractive index of rubidium vapor}
In order to characterize the optical response of rubidium vapor under laser excitation, the atom–field interaction is treated within the density matrix formalism and described by the optical Bloch equations.  
This approach relates the microscopic atomic dynamics to the macroscopic susceptibility, providing a framework for calculating the complex refractive index of the vapor.

A six-level system of $^{85}$Rb and $^{87}$Rb atoms is considered, which includes two ground-state hyperfine levels ($5S_{1/2}, F=2,3$ for $^{85}$Rb, and $F=1,2$ for $^{87}$Rb) and four excited-state hyperfine levels ($5P_{3/2}, F'=1,2,3,4$ for $^{85}$Rb, and $F'=0,1,2,3$ for $^{87}$Rb).
The optical Bloch equations of the system are given by:
\begin{equation}
	\begin{split}
		\frac{d\rho_{nm}}{dt} =&\; \frac{-i}{\hbar}\:[\widehat H,\widehat\rho\:]_{nm}-\gamma_{nm}\rho_{nm},\qquad n\neq m, \\
		\frac{d\rho_{nn}}{dt} =&\; \frac{-i}{\hbar}\:[\widehat H,\widehat\rho\:]_{nn}+\sum\limits_{E_m>E_n}\Gamma_{nm}\rho_{mm}\\
		&-\sum\limits_{E_m<E_n}\Gamma_{mn}\rho_{nn} - \Gamma_t\rho_{nn} + \Gamma_{nn} ,
		\label{Eq:obe}
	\end{split}
\end{equation}
where $\rho_{nn}$ are hyperfine level populations and $\rho_{nm}=\rho_{mn}^*$ are coherences induced in the system. 
$n=1,2,...,6$ and $m=1,2,...,6$ denote hyperfine levels, where 1 marks the lowest energy level, and 6 marks the highest energy level. 
The Hamiltonian of the system is $\widehat H=\widehat H_0+\widehat H_{int}$, where $\widehat H_0$ is the Hamiltonian of the free atom, and $(\widehat H_{int})_{nm}=-\mu_{nm}E(t)$ represents the interaction of the atom with the laser electric field $E(t)=E_0e^{i\omega t}$, with $\mu_{nm}$ the transition dipole moment \cite{steck}.

Coherence damping constants are given by: 
\begin{equation}
	\gamma_{nm}=\frac{\Gamma_n+\Gamma_m}{2} + \Gamma_t + \gamma_{pb},
	\label{Eq:gammanm}
\end{equation}
where $\Gamma_n$ and $\Gamma_m$ are relaxation rates of levels $n$ and $m$, respectively, whereas $\Gamma_{nm}$ is population decay from level $m$ to level $n$ \cite{steck}. 
$\Gamma_t$ accounts for the loss of atoms from the interaction zone with the laser beam due to their thermal motion in the vapor. 
This term includes the transit time broadening effects into the model, where the broadening rate $\Gamma_t = 1/\tau_t$ is determined by the transit time $\tau_t$ of atoms through the laser beam. 
In typical Rb vapor cells that are few centimeters long $\tau_t=\bar{d}/\bar{v}$, where $\bar{v}$ is the mean two-dimensional velocity of atoms in the vapor and the average distance $\bar{d}$ is found by averaging all possible paths through a circle representing the laser beam cross section \cite{sagle1996}. 
$\bar{d}=\pi D/4$, where $D$ is the laser beam FWHM. 
Once out of the interaction zone with the laser beam, the atoms quickly spontaneously decay into the ground state. 
Thus only the ground-state populations are replenished as a result of thermal motion of atoms in the vapor ($\Gamma_{nn} = 0$ for the excited state), and the atoms entering the beam are in a statistical mixture of the two ground-state hyperfine levels, i.e. $\Gamma_{11}=\Gamma_t \: g_1/(g_1+g_2)$ and $\Gamma_{22}=\Gamma_t \: g_2/(g_1+g_2)$, where $g=2F+1$ (see Supporting Information for a more detailed analysis of $\Gamma_{nn}$, including the discussion on how spin-preserving experimental conditions, e.g. a paraffin coating of the vapor cell, can be included in the model). 

The last term in the coherence damping constants in Eq. \ref{Eq:gammanm}, $\gamma_{pb}$, accounts for the pressure broadening. 
It can originate from the presence of a buffer gas or from Rb-Rb self-broadening (self-broadening at vapor temperatures below 100 °C can usually be neglected) \cite{weller2011}.

The spin-exchange inelastic collisions are included in the model through the $\Gamma_{12}$ and $\Gamma_{21}$ rates that describe the transfer of population between the ground hyperfine levels. 
They can usually be neglected for vapor temperatures below 100 °C, as the spin-exchange rate is about 2 kHz for the vapor temperature of 120 °C \cite{gibbs1967}. 
Population transfer between the excited hyperfine levels induced by spin-exchange inelastic collisions is neglected as it is typically about three orders of magnitude smaller than the spontaneous relaxation rates. 

In order to account for longitudinal atomic motion in the Doppler-broadened Rb vapor, Eq. \ref{Eq:obe} is solved for discrete velocity groups and subsequently averaged over the complete longitudinal (one-dimensional) Maxwell–Boltzmann atomic velocity distribution.

Eq. \ref{Eq:obe} is solved using the rotating wave approximation, and represents a system of 36 coupled differential equations that define the time evolution of the density matrix elements that describe the atomic response of six-level Rb atoms to laser light. 
We use the steady state solution of Eq. \ref{Eq:obe} to calculate the refractive index. 
Of particular interest is the steady state solution for the $\rho_{n1}=\sigma_{n1}e^{-i\omega t}$ coherence, where $n=3,4,5$, i.e. the coherence induced by the laser field between the levels 5S$_{1/2}, F=2$ and 5P$_{3/2}, F'=1,2,3$ in $^{85}$Rb (5S$_{1/2}, F=1$ and 5P$_{3/2}, F'=0,1,2$ in $^{87}$Rb). 
$\sigma_{n1}$ is the slowly varying envelope of the coherence and defines the system response to the laser light regarding transitions from the lower ground-state hyperfine level. 
Similarly, of interest is the steady state solution of Eq. \ref{Eq:obe} for the $\rho_{m2}=\sigma_{m2}e^{-i\omega t}$ coherence, where $m=4,5,6$, i.e. the coherence induced by the laser field between the levels 5S$_{1/2}, F=3$ and 5P$_{3/2}, F'=2,3,4$ in $^{85}$Rb (5S$_{1/2}, F=2$ and 5P$_{3/2}, F'=1,2,3$ in $^{87}$Rb), and $\sigma_{m2}$ defines the system response to the laser light regarding transitions from the upper ground-state hyperfine level.

The polarization of the medium induced by the probe laser is
\begin{equation}
	P=N\sum\limits_{n}\mu_{1n}\rho_{n1} + N\sum\limits_{m}\mu_{2m}\rho_{m2},
	\label{Eq:polarisation}
\end{equation}
where $N$ is the concentration of atoms, $n=3,4,5$, and $m=4,5,6$. 
The susceptibility of atoms to the probe laser electric field is then given by:
\begin{equation}
	\chi=\frac{N}{\epsilon_0 E_0}\left( \sum\limits_{n}\mu_{1n}\sigma_{n1} + \sum\limits_{m}\mu_{2m}\sigma_{m2} \right).
	\label{Eq:chi}
\end{equation}
Using $n=\sqrt{1+\chi}\approx 1+\chi/2$, the refractive index can be written as
\begin{equation}
	n=1+n_{Re}+in_{Im},
	\label{Eq:n}
\end{equation}
where the real and imaginary parts of the refractive index are given by
\begin{equation}
	n_{Re}=\frac{N}{2\epsilon_0 E_0} \left[ \sum\limits_{n}\mu_{1n} \operatorname{Re}(\sigma_{n1}) + \sum\limits_{m}\mu_{2m} \operatorname{Re}(\sigma_{m2}) \right],
	\label{Eq:Ren}
\end{equation}
\begin{equation}
	n_{Im}=\frac{N}{2\epsilon_0 E_0} \left[ \sum\limits_{n}\mu_{1n} \operatorname{Im}(\sigma_{n1}) + \sum\limits_{m}\mu_{2m} \operatorname{Im}(\sigma_{m2}) \right].
	\label{Eq:Imn}
\end{equation}
Eqs. \ref{Eq:Ren} and \ref{Eq:Imn} define the real and imaginary part of the complex refractive index of rubidium vapor based on the steady state solution of Eq. \ref{Eq:obe}. 

In order to theoretically model the effects related to the propagation of laser light through the rubidium vapor, we solve the nonlinear paraxial wave equation:
\begin{equation}
	\frac{\partial E_0}{\partial z} = \frac{i}{2k}\left( \frac{\partial^2}{\partial x^2} + \frac{\partial^2}{\partial y^2} \right)E_0 + ik\left(n_{Re}+in_{Im}\right)E_0,
	\label{Eq:paraxial}
\end{equation}
where $k$ is the laser wavevector, and $E_0$, $n_{Re}$ and $n_{Im}$ depend on the transversal coordinates $x$ and $y$. 
Eq. \ref{Eq:paraxial} is solved using the split-step Fourier method \cite{agrawal} to obtain a spatially varying phase that is accumulated by the laser beam as a result of propagation through the Rb vapor. 
Nonlinear phase simulations are implemented as follows: first Eq. \ref{Eq:obe} is solved parametrically in $E_0$, i.e. for a range of $E_0$ values, and the corresponding complex refractive index is calculated using Eqs. \ref{Eq:Ren} and \ref{Eq:Imn}. 
Then Eq. \ref{Eq:paraxial} is solved, where in each propagation step $n_{Re}(x,y)$ and $n_{Im}(x,y)$ are evaluated by interpolation using $E_0(x,y)$, and the results of Eqs. \ref{Eq:Ren} and \ref{Eq:Imn}. 
We note that $n_{Re}$ and $n_{Im}$ are in general nonlinear in $E_0$, and thus Eq. \ref{Eq:paraxial} is also nonlinear in $E_0$.

%%%%%%%%%%%%%%%%%%%%%%%%%%%%%%%%%%%%%%%%%%%%%%%%%%%%%%%%%%%%%%%%%%%%%%%%%%%%%%%%%%%%%%%%%%%%%%%%%%%%%%%%%%%%%%%%%%%%%%
\section{Experiment}
Interferometric techniques allow for the measurement of the local phase difference between a reference laser beam and a beam that has interacted with a nonlinear medium. 
In our experiment this is achieved by placing a rubidium vapor cell in one arm of a Mach-Zehnder interferometer. 
The two beams originating from the different interferometer arms are recombined on a camera at a small angle $\theta$, creating a spatial interference fringe pattern. 
The probe beam of 1.7 mm diameter at $1/e^2$ intensity was used. 

\begin{figure}[h]
	\centering
	\includegraphics[width=1\linewidth]{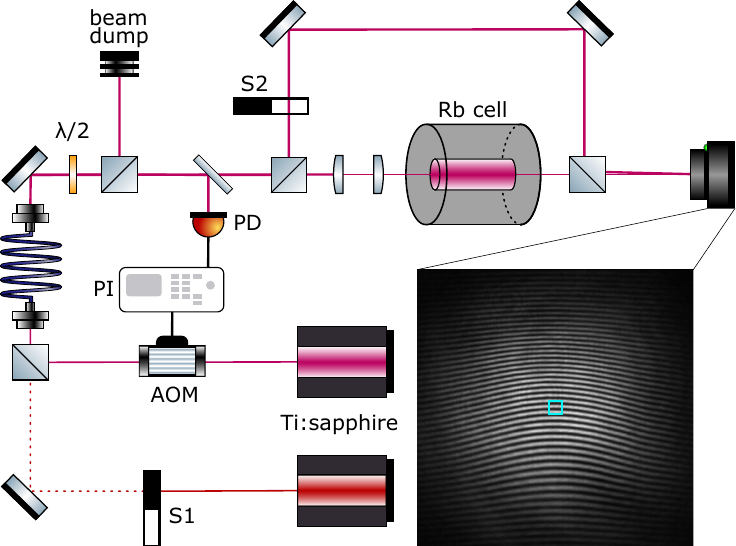}
	\caption{Scheme of the experimental setup for the interferometric measurement of the refractive index of Rb vapor. The resulting interferogram is shown in the bottom right with the small square indicating the region used for calculating the phase.}
	\label{fig:setup}
\end{figure}

A 5 cm glass cell containing isotopically pure $^{87}$Rb vapor was used.
To reduce stray magnetic fields, the cell was housed inside a $\mu$-metal shield, while the heating element was kept outside. 
Heat transfer to the cell was achieved using heat pipes, and the temperature was stabilized to about 1 °C. 

A wavemeter (High Finesse Ångstrom WS/7) and an acousto-optic modulator (AOM) were used to stabilize the frequency and power of the probe laser (Sirah Matisse CR). 
An additional frequency-stabilized laser beam (Sirah Matisse CR) was also sent through the Rb vapor, and was held at a constant power and frequency, far detuned from the Rb resonances. 
It is used to record phase variations caused by ambient air temperature fluctuations and other environmental conditions, which are later subtracted from the phase measured by the probe laser. 

The interference pattern is recorded using a camera (Flir Blackfly S BFS-U3-120S4M). 
Changes in the phase of the probe laser beam (linear phase) cause the interference fringes to shift. 
A spatially varying phase (nonlinear phase) on the other hand causes the interference fringes to bend, as illustrated in Fig. \ref{fig:setup}. 
In both regimes we select a small region in the center of the interference image (indicated by a small square in Fig. \ref{fig:setup}), containing approximately two interference fringes, to calculate the phase. 
The intensity in this selected region is first integrated along the axis parallel to the fringes, resulting in a space-averaged one-dimensional (1D) intensity.
The phase is extracted by fitting the sinusoidal function to the space-averaged 1D intensity.
During measurements, care is taken to vary experimental parameters in small increments, ensuring the phase change between consecutive data points remains below $\pi$. 
Finally, a phase-unwrapping algorithm is applied to the fitted phase values, and the phase related to the environmental noise is subtracted. 
A similar interferometric technique was also used in Ref. \cite{aladjidi} to study the transit effects for non-linear refractive index measurement in warm atomic vapors. 

In the linear refractive index regime, the phase $\delta\phi$ accumulated by the laser beam after propagating through the vapor is directly related to the vapor refractive index by the relation $\delta\phi=k\:\delta n\:L$, where $L$ is the vapor length.
We note that when measuring phase as a function of frequency one must include an additional linear phase shift that is related to the change of $k$ with frequency, and which is due to the difference in path lengths between the two arms of the interferometer. 
This background contribution was experimentally characterized by measuring the phase as a function of frequency far from the Rb resonance lines. 

In the nonlinear regime, the relation between $\delta\phi$ and $\delta n$ is not straightforward, and in general one needs to solve the nonlinear paraxial wave equation (Eq. \ref{Eq:paraxial}) to obtain the accumulated nonlinear phase from the nonlinear refractive index. 
Experimentally, nonlinear phase is obtained by measuring the phase as a function of the probe laser power.

%%%%%%%%%%%%%%%%%%%%%%%%%%%%%%%%%%%%%%%%%%%%%%%%%%%%%%%%%%%%%%%%%%%%%%%%%%%%%%%%%%%%%%%%%%%%%%%%%%%%%%%%%%%%%%%%%%%%%%
\section{Results}
\begin{figure}[h]
	\centering
	\includegraphics[width=1\linewidth,trim={2.cm 1.0cm 3.cm 1.0cm},clip]{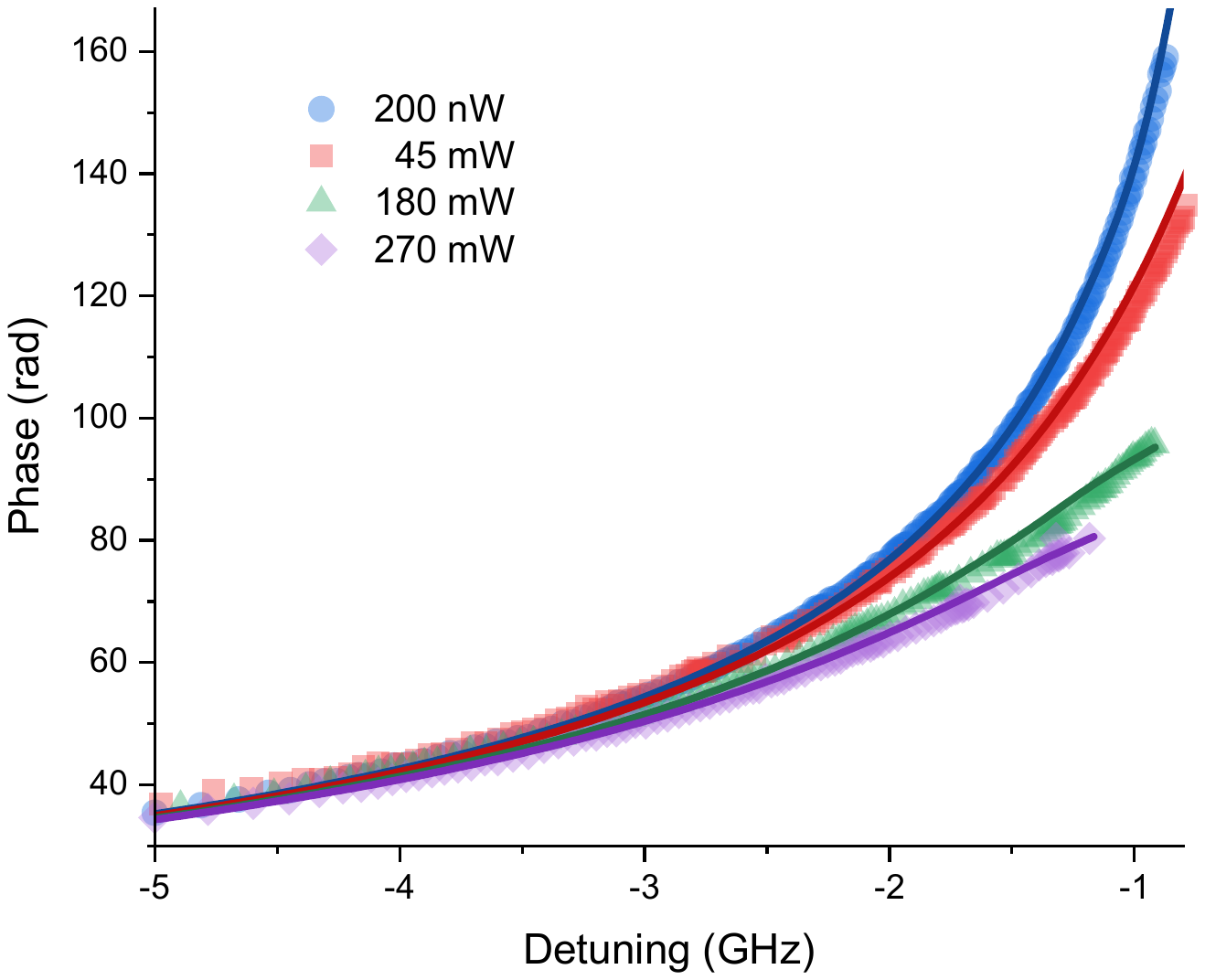}
	\caption{Phase as a function of detuning from the $^{87}$Rb 5S$_{1/2}, F=2 \rightarrow$ 5P$_{3/2}, F'=2$ transition. Measured results (data points) and theoretical simulations (lines) are shown for different laser power. $T=119$°C, $N=1.54\times10^{13}$ cm$^{-3}$.}
	\label{fig:phase_red}
\end{figure}

\begin{figure}[h]
	\centering
	\includegraphics[width=1\linewidth,trim={2.cm 1.0cm 3.cm 1.0cm},clip]{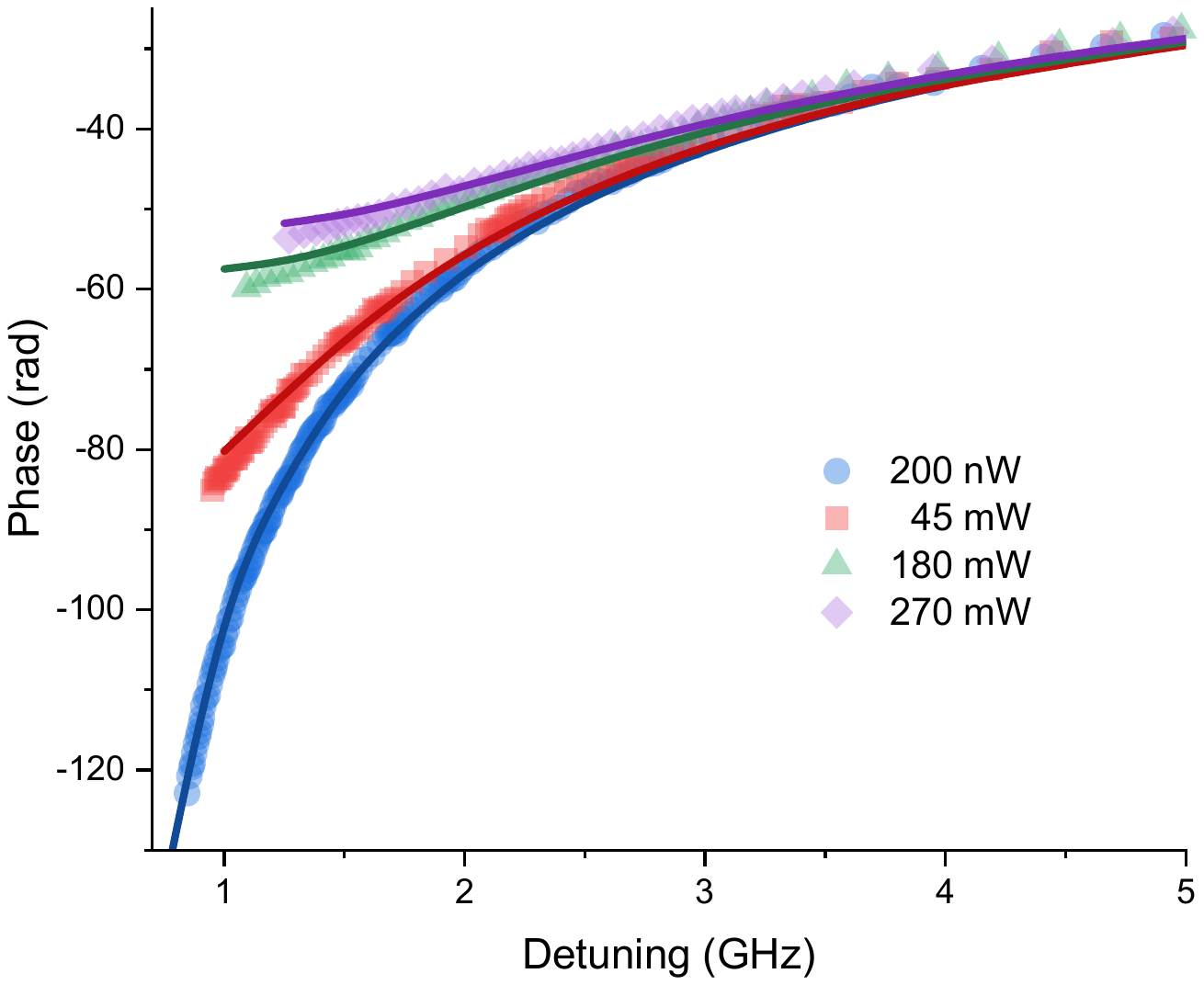}
	\caption{Phase as a function of detuning from the $^{87}$Rb 5S$_{1/2}, F=1 \rightarrow$ 5P$_{3/2}, F'=2$ transition. Measured results (data points) and theoretical simulations (lines) are shown for different laser power. Cell temperature $T$ and atom concentration $N$ are the same as in Fig. \ref{fig:phase_red}.}
	\label{fig:phase_blue}
\end{figure}

We start our investigation of the refractive index of Rb vapor by measuring the phase accumulated by the probe laser after propagating through the Rb vapor as a function of probe laser frequency. 
Both experimental and theoretical results are shown in Fig. \ref{fig:phase_red} for the red detuning from the $^{87}$Rb 5S$_{1/2}, F=2 \rightarrow$ 5P$_{3/2}, F'=1,2,3$ hyperfine line, and in Fig. \ref{fig:phase_blue} for the blue detuning from the $^{87}$Rb 5S$_{1/2}, F=1 \rightarrow$ 5P$_{3/2}, F'=0,1,2$ line. 
The Rb vapor cell used in the experiment is an isotopically pure $^{87}$Rb vapor cell. 
Still, we experimentally observe the presence of $^{85}$Rb isotope of about 2$\%$ in the vapor. 
This small but noticeable presence of $^{85}$Rb affects the refractive index in the blue wing of the $^{87}$Rb 5S$_{1/2}, F=2 \rightarrow$ 5P$_{3/2}, F'=1,2,3$ line, and in the red wing of the  $^{87}$Rb 5S$_{1/2}, F=1 \rightarrow$ 5P$_{3/2}, F'=0,1,2$ line (see Supporting Information for an illustration of $^{85}$Rb and $^{87}$Rb hyperfine line interrelation), making the analysis of the refractive index in these ranges not straightforward. 
We thus focus our investigation on the red detuning from the $^{87}$Rb 5S$_{1/2}, F=2 \rightarrow$ 5P$_{3/2}, F'=1,2,3$ hyperfine line, and on the blue detuning from the $^{87}$Rb 5S$_{1/2}, F=1 \rightarrow$ 5P$_{3/2}, F'=0,1,2$ line, and call them the red and blue detuning throughout the paper, respectively. 

Phase measurements for four different probe laser powers are shown in Figs. \ref{fig:phase_red} and \ref{fig:phase_blue}. 
For detunings close to resonance ($<1$ GHz) the probe laser is strongly absorbed thus obstructing the phase measurement. 
The lowest power of 200 nW corresponds to the linear regime, i.e. the refractive index does not change with laser power at this power level. 
Theoretical calculations show excellent agreement with the measured results in the linear regime. 
This is further confirmed by the comparison of theoretical simulations with experimental results for the hyperfine absorption spectra at different vapor temperatures from Ref. \cite{siddons2008} (see Supporting Information). 

Increasing the probe laser power introduces the nonlinear effects in the interaction of laser light with Rb atoms. 
This is observed as the reduction of the phase values for both the red and blue detuning towards zero, corresponding to the negative nonlinear refractive index for the red detuning, and positive nonlinear refractive index for the blue detuning. 
In general, very good quantitative agreement between the theoretical and experimental results for the accumulated phase is obtained. 

Positive nonlinear refractive index for the blue detuning leads to self-focusing of the laser beam. 
This makes a detailed analysis of the frequency and power dependence of the nonlinear refractive index impractical for the blue detuning, since the beam starts to distort as soon as the nonlinearity becomes noticeable. 
The beam distortions result from small initial beam intensity irregularities that are strongly emphasized through nonlinear propagation under the self-focusing conditions and eventually hinder the accurate phase measurement. 
We thus focus our investigation on the red detuning, where self-defocusing of the probe beam is observed, and nonlinear phase can readily be measured in a broad power and frequency range. 

\begin{figure}[h]
	\centering
	\includegraphics[width=1\linewidth,trim={1.5cm 1.0cm 3.cm 2.0cm},clip]{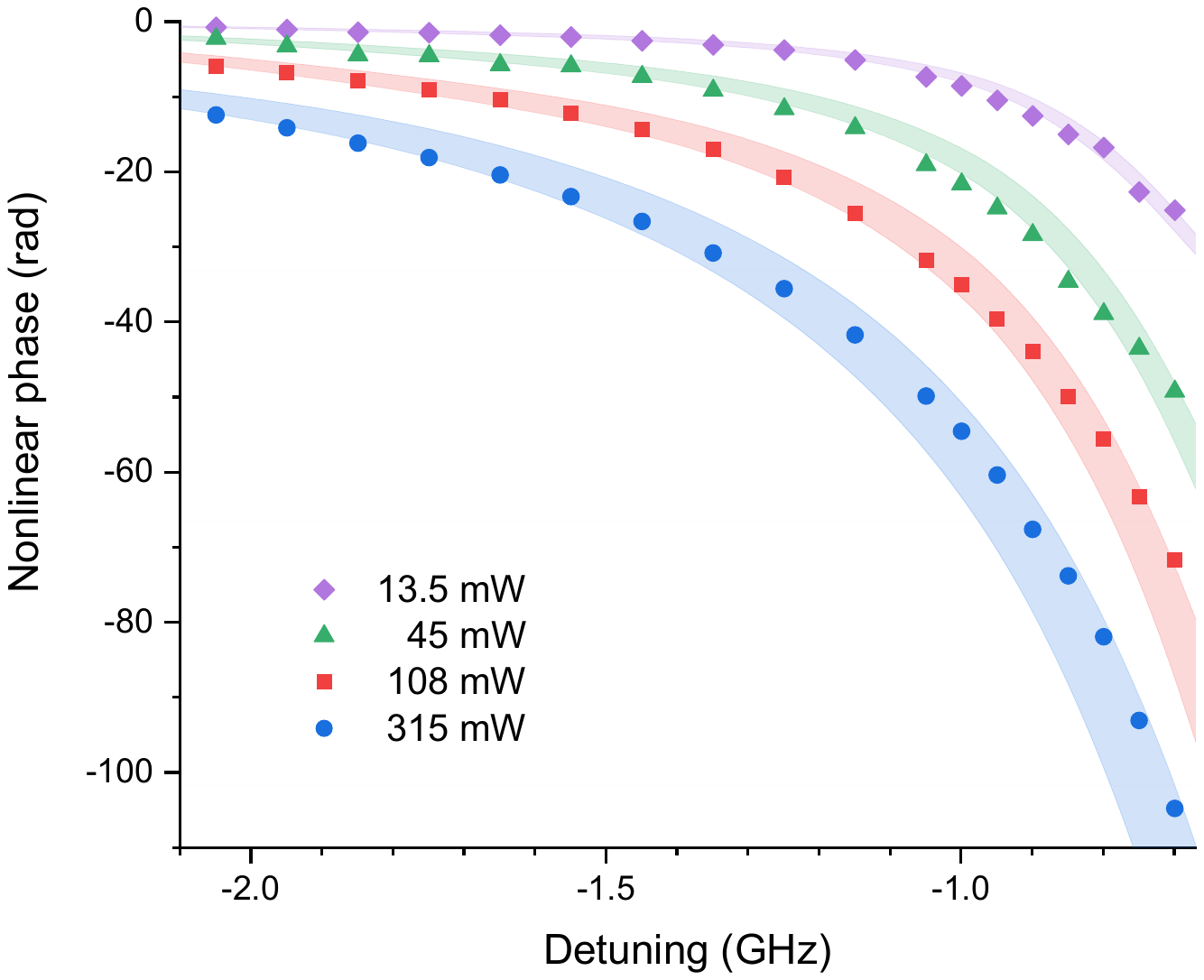}
	\caption{Nonlinear phase as a function of detuning from the $^{87}$Rb 5S$_{1/2}, F=2 \rightarrow$ 5P$_{3/2}, F'=2$ transition. Measured results (data points) and theoretical simulations (shaded areas) are shown for different laser power, where the shaded areas represent the uncertainty of the calculated nonlinear phase resulting from the uncertainty of the concentration of $^{87}$Rb atoms.}
	\label{fig:nlin_phase_fscan}
\end{figure}

The nonlinear phase as a function of detuning is shown in Fig. \ref{fig:nlin_phase_fscan} for different laser powers. 
The nonlinear phase strongly depends on the concentration of atoms. 
For the theoretical calculations of the nonlinear phase accumulated by the probe laser after passing through the vapor we use $N=1.37\pm0.17\times10^{13}$ cm$^{-3}$, which is determined from the measurement of the probe laser transmission through the vapor. 
This experimental uncertainty in the Rb concentration of about $\pm12\%$ is illustrated in Fig. \ref{fig:nlin_phase_fscan} as a coloured shaded region, where the lower line for a given color (detuning) corresponds to the calculated nonlinear phase for $N=1.21\times10^{13}$ cm$^{-3}$, whereas the upper line corresponds to $N=1.54\times10^{13}$ cm$^{-3}$. 
The quantitative agreement between the measured and calculated nonlinear phase frequency dependence is very good. 
A minor systematic deviation is observed as the calculated nonlinear phase values for smaller detunings tend to be slightly higher than the measured values. 
Nevertheless, the overall agreement is very good, particularly taking into account that an ab-initio theoretical model is used with no free parameters, where all parameters entering the model are determined by the experimental values related to the laser beam and the Rb vapor. 

\begin{figure}[h]
	\centering
	\includegraphics[width=1\linewidth,trim={1.5cm 1.0cm 3.cm 2.0cm},clip]{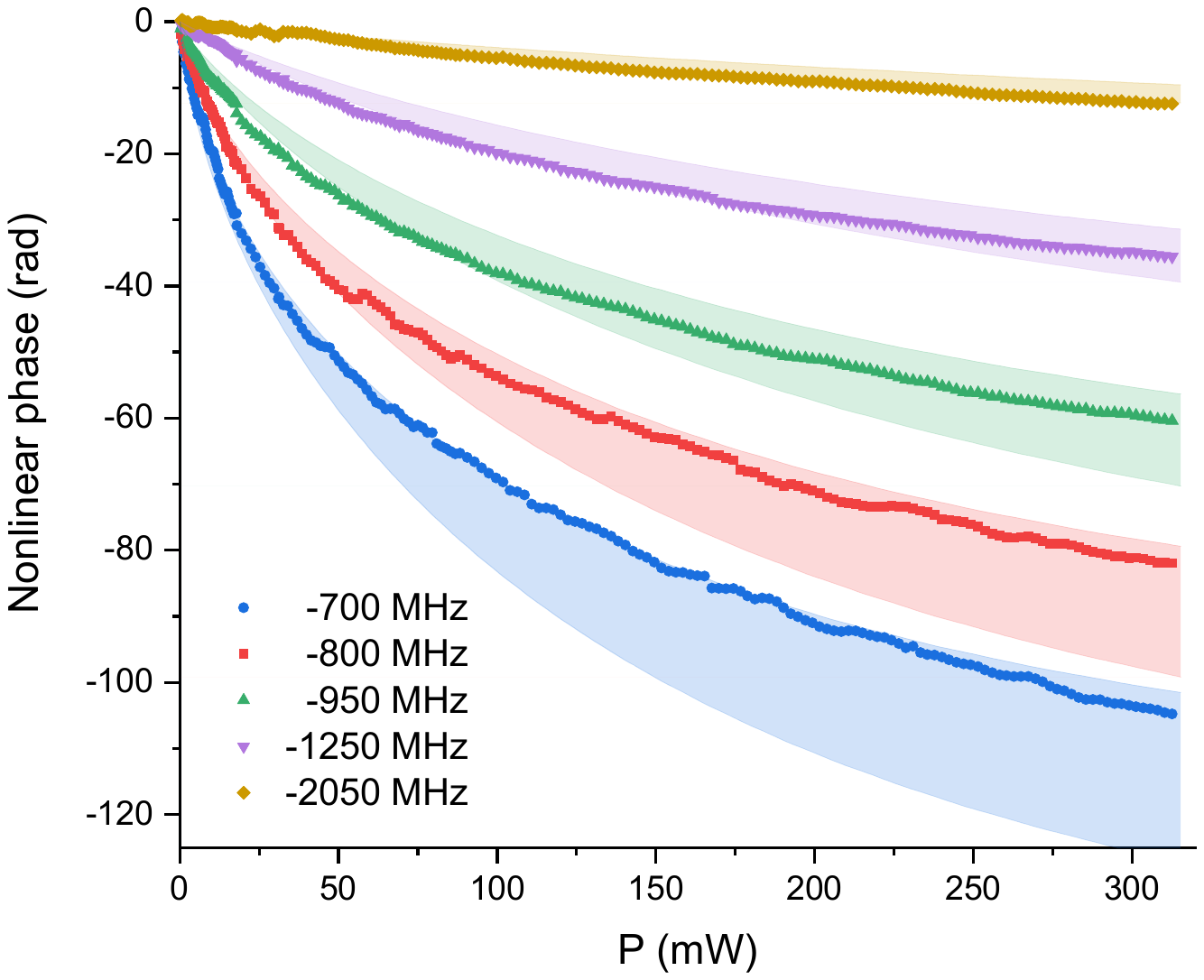}
	\caption{Nonlinear phase as a function of laser power. Measured results (data points) and theoretical simulations (shaded areas) are shown for different laser detunings from the $^{87}$Rb 5S$_{1/2}, F=2 \rightarrow$ 5P$_{3/2}, F'=2$ transition, where the shaded areas represent the uncertainty of the calculated nonlinear phase resulting from the uncertainty of the concentration of $^{87}$Rb atoms.}
	\label{fig:nlin_phase_pscan}
\end{figure}

The nonlinear phase as a function of laser power is shown in Fig. \ref{fig:nlin_phase_pscan} for different laser detunings. 
A clear saturation behavior of the nonlinear phase is observed that is indicative of the saturable nonlinear refractive index of Rb vapor. 
We note that, unlike the linear regime where phase is directly proportional to the refractive index, the relation of the nonlinear phase and the nonlinear refractive index is more complex. 
In general, one must calculate the nonlinear laser beam propagation using Eq. \ref{Eq:paraxial} in order to calculate the nonlinear phase that is accumulated by the laser beam as a result of the propagation in the medium characterized by a nonlinear refractive index $n=n_\textnormal{L}+n_\textnormal{{NL}}$, where $n_\textnormal{L}$ and $n_\textnormal{NL}$ are the linear and nonlinear part of the complex refractive index. 

\begin{figure}[h]
	\centering
	\includegraphics[width=1\linewidth,trim={1.5cm 1.0cm 3.cm 1.5cm},clip]{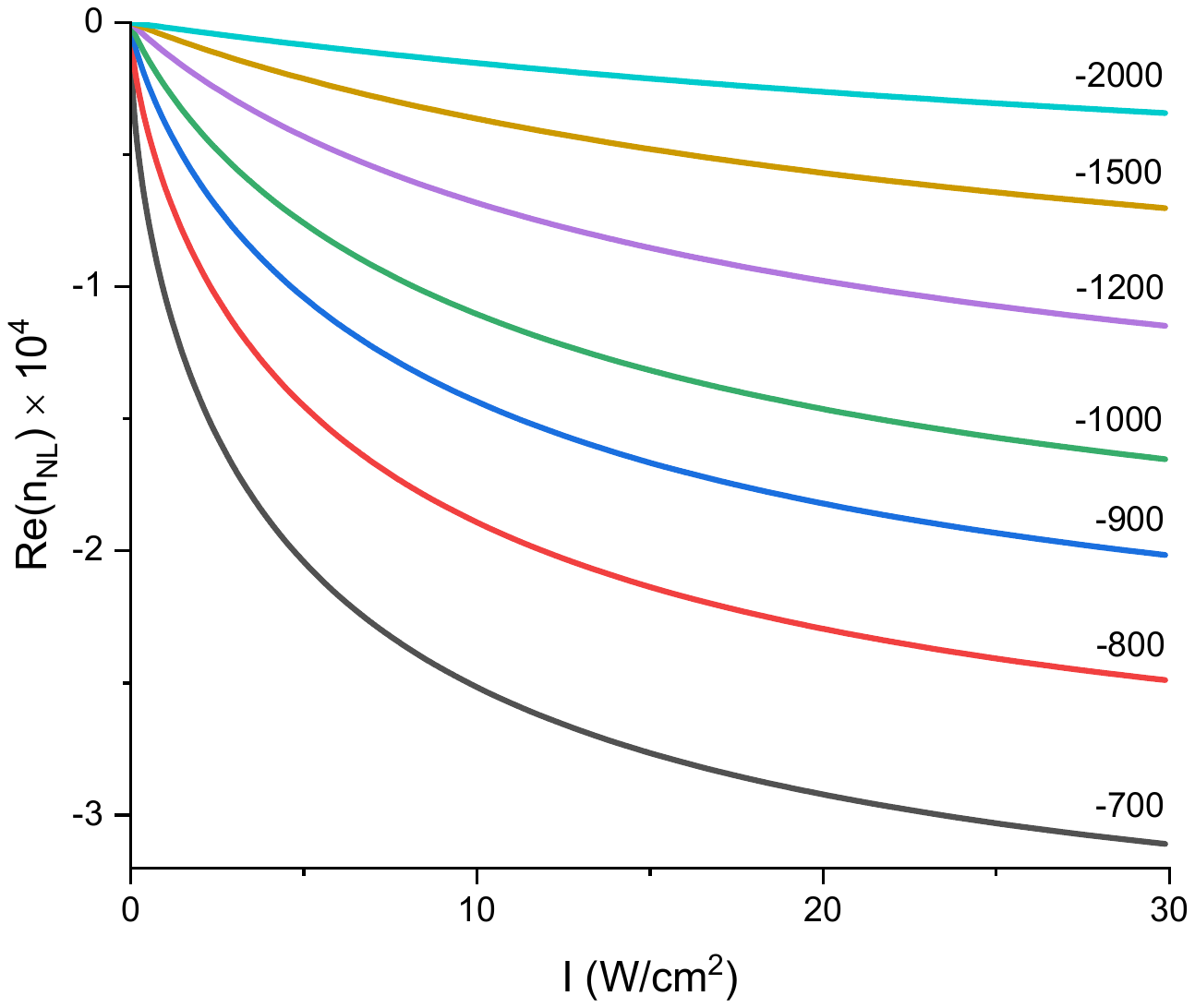}
	\caption{Calculated real part of the nonlinear refractive index $n_\textnormal{{NL}}$ as a function of laser intensity for different laser detunings from the $^{87}$Rb 5S$_{1/2}, F=2 \rightarrow$ 5P$_{3/2}, F'=2$ transition (indicated in MHz). The parameters used in the calculations are the same as in Figs. \ref{fig:nlin_phase_fscan} and \ref{fig:nlin_phase_pscan}, and $N=1.37\times10^{13}$ cm$^{-3}$.}
	\label{fig:n_nonlin_pscan}
\end{figure}

We show in Fig. \ref{fig:n_nonlin_pscan} the real part of the calculated nonlinear refractive index of $^{87}$Rb vapor for the same experimental parameters as in Figs. \ref{fig:nlin_phase_fscan} and \ref{fig:nlin_phase_pscan}, and $N=1.37\times10^{13}$ cm$^{-3}$. 
As seen in Fig. \ref{fig:n_nonlin_pscan}, nonlinear refractive index of up to $-3\times10^{-4}$ is obtained in our experiment. 

\begin{figure}[h]
	\centering
	\includegraphics[width=1\linewidth,trim={1.5cm 1.0cm 0.5cm 1.5cm},clip]{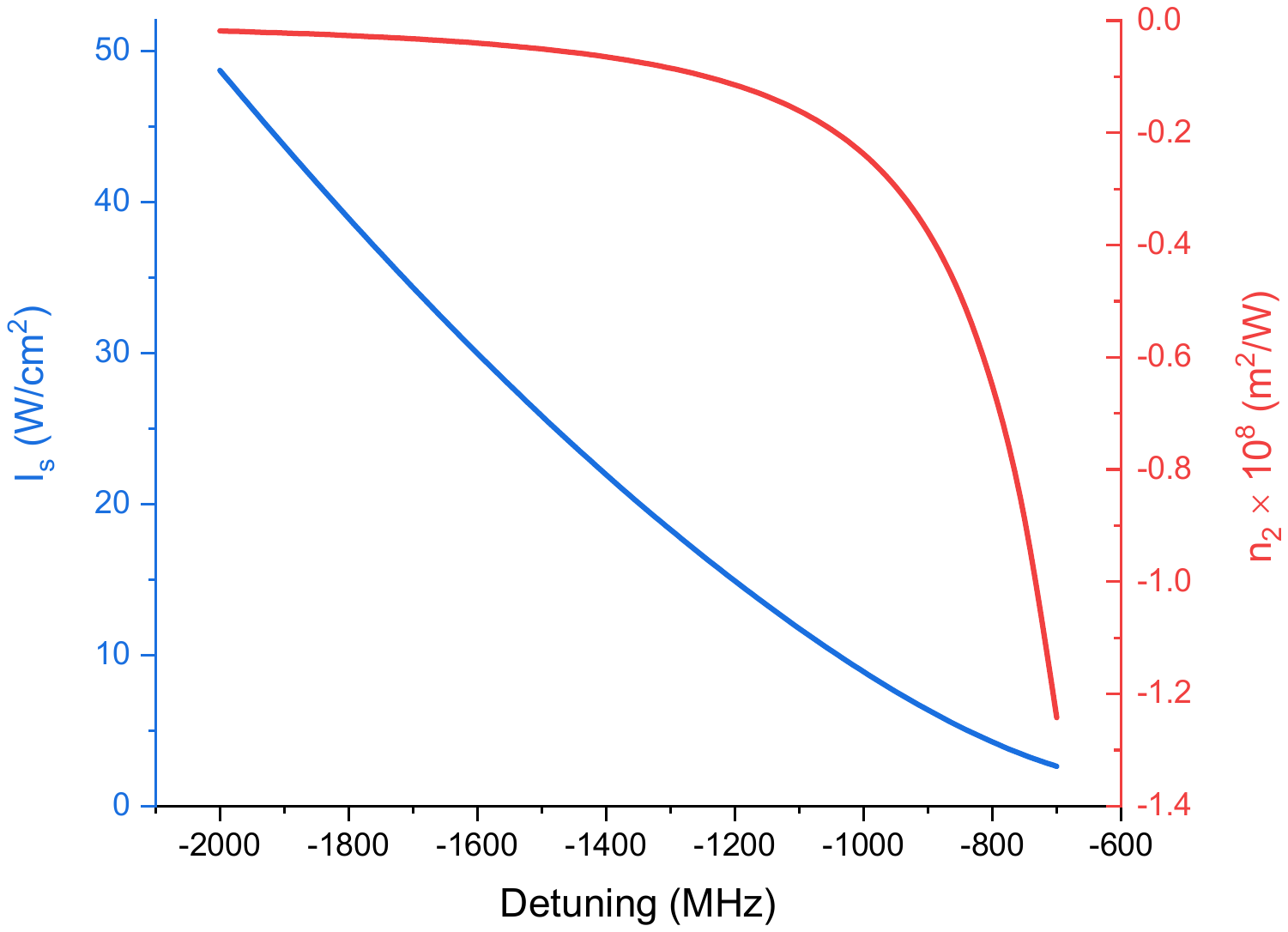}
	\caption{Unsaturated Kerr coefficient $n_2$ (red) and nonlinear refractive index saturation intensity $I_s$ (blue) obtained by fitting the Eq. \ref{Eq:n2} to the results shown in Fig. \ref{fig:n_nonlin_pscan} as a function of detuning from the $^{87}$Rb 5S$_{1/2}, F=2 \rightarrow$ 5P$_{3/2}, F'=2$ transition.}
	\label{fig:n2_Isat}
\end{figure}

As the nonlinear refractive index shows clear saturation behavior with laser intensity, we can fit Eq. \ref{Eq:n2} to the calculated Re$(n_\textnormal{{NL}})$ for different detunings. 
The resulting fitted parameters $n_2$ and $I_s$ are shown in Fig. \ref{fig:n2_Isat} as a function of detuning. 
$n_2$ represents the unsaturated nonlinear Kerr coefficient that defines the nonlinear refractive index of Rb vapor in the low intensity limit ($I\ll I_s$), in which Eq. \ref{Eq:n2} reduces to $n(I)=n_0+n_2I$. 
$I_s$ defines the saturation bahavior of Re$(n_\textnormal{{NL}})$. 
As seen in Fig. \ref{fig:n2_Isat}, up to only a few W/cm$^2$ are sufficient to saturate the nonlinear refractive index of warm $^{87}$Rb vapor for detunings below 1000 MHz. 
For example, only a few tens of milliwatts are sufficient for a millimeter size laser beam to saturate the nonlinear refractive index of $^{87}$Rb vapor at 117 °C and 700 MHz detuning. 

Nonlinear Kerr coefficient of up to $-1.2\times10^{-4}$ cm$^2$/W is obtained for 700 MHz detuning in our experimental conditions of $^{87}$Rb vapor with concentration $N=1.37\times10^{13}$ cm$^{-3}$. 
This is much larger than reported in Ref. \cite{mccormick2004}, where $n_2\approx10^{-7}$ cm$^2$/W for 1 GHz detunings from the D2 line in natural-abundance Rb vapor at 78 °C is reported. 
It is also larger than reported in Ref. \cite{wang2020}, where $n_2\approx10^{-6}$ cm$^2$/W for similar detunings in $^{87}$Rb vapor with concentration $N=1.55\times10^{12}$ cm$^{-3}$ is reported. 
This apparent discrepancy in $n_2$ is however easy to explain when the quoted $n_2$ values are scaled by the concentration of $^{87}$Rb atoms in these experiments. 

Finally, we note that both qualitatively and quantitatively similar behavior of $n_2$ and $I_s$ is obtained for the case of blue detuning (see Supporting Information).

%%%%%%%%%%%%%%%%%%%%%%%%%%%%%%%%%%%%%%%%%%%%%%%%%%%%%%%%%%%%%%%%%%%%%%%%%%%%%%%%%%%%%%%%%%%%%%%%%%%%%%%%%%%%%%%%%%%%%%
\section{Conclusion}
We have calculated the refractive index of warm rubidium vapor using a theoretical model based on optical Bloch equations for 6-level Rb atoms interacting with a probe laser. 
The model includes Doppler broadening, transit time broadening, pressure broadening, saturation, optical pumping, and spin-exchange inelastic collisions. 
Based on the calculated nonlinear refractive index, and using nonlinear paraxial wave equation, we have simulated the spatially varying nonlinear phase accumulated by the probe laser beam after passing through the $^{87}$Rb vapor. 
This theoretically simulated phase was then compared to the experimental phase obtained by using an interferometric technique that places a rubidium vapor cell in one arm of a Mach-Zehnder interferometer to create a spatial interference fringe pattern that enables measurement of the local phase difference between the reference beam and the beam that has interacted with the Rb vapor. 
We show very good quantitative agreement of the theoretical and experimental results. 

We provide Python scripts for all theoretical calculations presented in this work (\url{https://github.com/damiraumiler/Rb_D2_nonlinear_refractive_index}), including the refractive index calculation, that can readily be used in practical implementations for real-time simulations of the linear and nonlinear refractive index of Rb vapor.

\section{Acknowledgements}
N.Š. and D.A. acknowledge support from the project “Implementation of cutting-edge research and its application as part of the Scientific Center of Excellence for Quantum and Complex Systems, and Representations of Lie Algebras”, Grant No. PK.1.1.10.0004, co-financed by the European Union through the European Regional Development Fund - Competitiveness and Cohesion Programme 2021-2027.
We also acknowledge support from the project "Centre for Advanced Laser Techniques" (CALT), cofunded by the European Union through the European Regional Development Fund under the Competitiveness and Cohesion Operational Programme (Grant No. KK.01.1.1.05.0001).

\bibliography{n2_reference}

\begin{thebibliography}{19}
\expandafter\ifx\csname natexlab\endcsname\relax\def\natexlab#1{#1}\fi
\expandafter\ifx\csname bibnamefont\endcsname\relax
  \def\bibnamefont#1{#1}\fi
\expandafter\ifx\csname bibfnamefont\endcsname\relax
  \def\bibfnamefont#1{#1}\fi
\expandafter\ifx\csname citenamefont\endcsname\relax
  \def\citenamefont#1{#1}\fi
\expandafter\ifx\csname url\endcsname\relax
  \def\url#1{\texttt{#1}}\fi
\expandafter\ifx\csname urlprefix\endcsname\relax\def\urlprefix{URL }\fi
\providecommand{\bibinfo}[2]{#2}
\providecommand{\eprint}[2][]{\url{#2}}

\bibitem[{\citenamefont{Kitching}(2018)}]{kitching2018}
\bibinfo{author}{\bibfnamefont{J.}~\bibnamefont{Kitching}},
  \bibinfo{journal}{Appl. Phys. Rev.} \textbf{\bibinfo{volume}{5}},
  \bibinfo{pages}{031302} (\bibinfo{year}{2018}).

\bibitem[{\citenamefont{Ries et~al.}(2003)\citenamefont{Ries, Brezger, and
  Lvovsky}}]{ries2003}
\bibinfo{author}{\bibfnamefont{J.}~\bibnamefont{Ries}},
  \bibinfo{author}{\bibfnamefont{B.}~\bibnamefont{Brezger}}, \bibnamefont{and}
  \bibinfo{author}{\bibfnamefont{A.~I.} \bibnamefont{Lvovsky}},
  \bibinfo{journal}{Phys. Rev. A} \textbf{\bibinfo{volume}{68}},
  \bibinfo{pages}{025801} (\bibinfo{year}{2003}).

\bibitem[{\citenamefont{Mottola et~al.}(2023)\citenamefont{Mottola, Buser, and
  Treutlein}}]{treutlein2023}
\bibinfo{author}{\bibfnamefont{R.}~\bibnamefont{Mottola}},
  \bibinfo{author}{\bibfnamefont{G.}~\bibnamefont{Buser}}, \bibnamefont{and}
  \bibinfo{author}{\bibfnamefont{P.}~\bibnamefont{Treutlein}},
  \bibinfo{journal}{Phys. Rev. Lett.} \textbf{\bibinfo{volume}{131}},
  \bibinfo{pages}{260801} (\bibinfo{year}{2023}).

\bibitem[{\citenamefont{Šantić et~al.}(2018)\citenamefont{Šantić, Fusaro,
  Salem, Garnier, Picozzi, and Kaiser}}]{santic2018}
\bibinfo{author}{\bibfnamefont{N.}~\bibnamefont{Šantić}},
  \bibinfo{author}{\bibfnamefont{A.}~\bibnamefont{Fusaro}},
  \bibinfo{author}{\bibfnamefont{S.}~\bibnamefont{Salem}},
  \bibinfo{author}{\bibfnamefont{J.}~\bibnamefont{Garnier}},
  \bibinfo{author}{\bibfnamefont{A.}~\bibnamefont{Picozzi}}, \bibnamefont{and}
  \bibinfo{author}{\bibfnamefont{R.}~\bibnamefont{Kaiser}},
  \bibinfo{journal}{Phys. Rev. Lett.} \textbf{\bibinfo{volume}{120}},
  \bibinfo{pages}{055301} (\bibinfo{year}{2018}).

\bibitem[{\citenamefont{Glorieux et~al.}(2023)\citenamefont{Glorieux, Aladjidi,
  Lett, and Kaiser}}]{glorieux2023}
\bibinfo{author}{\bibfnamefont{Q.}~\bibnamefont{Glorieux}},
  \bibinfo{author}{\bibfnamefont{T.}~\bibnamefont{Aladjidi}},
  \bibinfo{author}{\bibfnamefont{P.~D.} \bibnamefont{Lett}}, \bibnamefont{and}
  \bibinfo{author}{\bibfnamefont{R.}~\bibnamefont{Kaiser}},
  \bibinfo{journal}{New J. Phys.} \textbf{\bibinfo{volume}{25}},
  \bibinfo{pages}{051201} (\bibinfo{year}{2023}).

\bibitem[{\citenamefont{Vulić et~al.}(2026)\citenamefont{Vulić, Šantić,
  Buljan, and Aumiler}}]{vulic2026}
\bibinfo{author}{\bibfnamefont{V.}~\bibnamefont{Vulić}},
  \bibinfo{author}{\bibfnamefont{N.}~\bibnamefont{Šantić}},
  \bibinfo{author}{\bibfnamefont{H.}~\bibnamefont{Buljan}}, \bibnamefont{and}
  \bibinfo{author}{\bibfnamefont{D.}~\bibnamefont{Aumiler}},
  \bibinfo{journal}{to be published}  (\bibinfo{year}{2026}).

\bibitem[{\citenamefont{McCormick et~al.}(2004)\citenamefont{McCormick, Solli,
  and Chiao}}]{mccormick2004}
\bibinfo{author}{\bibfnamefont{C.~F.} \bibnamefont{McCormick}},
  \bibinfo{author}{\bibfnamefont{D.~R.} \bibnamefont{Solli}}, \bibnamefont{and}
  \bibinfo{author}{\bibfnamefont{R.~Y.} \bibnamefont{Chiao}},
  \bibinfo{journal}{Phys. Rev. A} \textbf{\bibinfo{volume}{69}},
  \bibinfo{pages}{023804} (\bibinfo{year}{2004}).

\bibitem[{\citenamefont{McCormick et~al.}(2003)\citenamefont{McCormick, Solli,
  Chiao, and Hickmann}}]{mccormick2003}
\bibinfo{author}{\bibfnamefont{C.~F.} \bibnamefont{McCormick}},
  \bibinfo{author}{\bibfnamefont{D.~R.} \bibnamefont{Solli}},
  \bibinfo{author}{\bibfnamefont{R.~Y.} \bibnamefont{Chiao}}, \bibnamefont{and}
  \bibinfo{author}{\bibfnamefont{J.~M.} \bibnamefont{Hickmann}},
  \bibinfo{journal}{J. Opt. Soc. Am. B} \textbf{\bibinfo{volume}{20}},
  \bibinfo{pages}{2480} (\bibinfo{year}{2003}).

\bibitem[{\citenamefont{J.Wu et~al.}(2022)\citenamefont{J.Wu, Jia, S.Wang,
  X.Wang, Yuan, L.Wang, Hu, Chen, and Xu}}]{wu2022}
\bibinfo{author}{\bibnamefont{J.Wu}},
  \bibinfo{author}{\bibfnamefont{P.}~\bibnamefont{Jia}},
  \bibinfo{author}{\bibnamefont{S.Wang}},
  \bibinfo{author}{\bibnamefont{X.Wang}},
  \bibinfo{author}{\bibfnamefont{J.}~\bibnamefont{Yuan}},
  \bibinfo{author}{\bibnamefont{L.Wang}},
  \bibinfo{author}{\bibfnamefont{Y.}~\bibnamefont{Hu}},
  \bibinfo{author}{\bibfnamefont{Z.}~\bibnamefont{Chen}}, \bibnamefont{and}
  \bibinfo{author}{\bibfnamefont{J.}~\bibnamefont{Xu}}, \bibinfo{journal}{Opt.
  Express} \textbf{\bibinfo{volume}{30}}, \bibinfo{pages}{43012}
  (\bibinfo{year}{2022}).

\bibitem[{\citenamefont{Wang et~al.}(2020)\citenamefont{Wang, Yuan, Wang, Xiao,
  and Jia}}]{wang2020}
\bibinfo{author}{\bibfnamefont{S.}~\bibnamefont{Wang}},
  \bibinfo{author}{\bibfnamefont{J.}~\bibnamefont{Yuan}},
  \bibinfo{author}{\bibfnamefont{L.}~\bibnamefont{Wang}},
  \bibinfo{author}{\bibfnamefont{L.}~\bibnamefont{Xiao}}, \bibnamefont{and}
  \bibinfo{author}{\bibfnamefont{S.}~\bibnamefont{Jia}}, \bibinfo{journal}{Opt.
  Express} \textbf{\bibinfo{volume}{28}}, \bibinfo{pages}{38334}
  (\bibinfo{year}{2020}).

\bibitem[{\citenamefont{Levine and Du}(2023)}]{levine2023}
\bibinfo{author}{\bibfnamefont{Z.~H.} \bibnamefont{Levine}} \bibnamefont{and}
  \bibinfo{author}{\bibfnamefont{Z.}~\bibnamefont{Du}}, \bibinfo{journal}{J.
  Opt. Soc. Am. B} \textbf{\bibinfo{volume}{40}}, \bibinfo{pages}{3190}
  (\bibinfo{year}{2023}).

\bibitem[{\citenamefont{Levine}(2025)}]{levine2025}
\bibinfo{author}{\bibfnamefont{Z.~H.} \bibnamefont{Levine}},
  \bibinfo{journal}{J. Opt. Soc. Am. B} \textbf{\bibinfo{volume}{42}},
  \bibinfo{pages}{1310} (\bibinfo{year}{2025}).

\bibitem[{\citenamefont{Steck}()}]{steck}
\bibinfo{author}{\bibfnamefont{D.}~\bibnamefont{Steck}},
  \emph{\bibinfo{title}{Alkali d line data}},
  \urlprefix\url{https://steck.us/alkalidata/}.

\bibitem[{\citenamefont{Sagle et~al.}(1996)\citenamefont{Sagle, Namiotka, and
  Huennekens}}]{sagle1996}
\bibinfo{author}{\bibfnamefont{J.}~\bibnamefont{Sagle}},
  \bibinfo{author}{\bibfnamefont{R.~K.} \bibnamefont{Namiotka}},
  \bibnamefont{and}
  \bibinfo{author}{\bibfnamefont{J.}~\bibnamefont{Huennekens}},
  \bibinfo{journal}{J. Phys. B: At. Mol. Opt. Phys.}
  \textbf{\bibinfo{volume}{29}}, \bibinfo{pages}{2629} (\bibinfo{year}{1996}).

\bibitem[{\citenamefont{Weller et~al.}(2011)\citenamefont{Weller, Bettles,
  Siddons, Adams, and Hughes}}]{weller2011}
\bibinfo{author}{\bibfnamefont{L.}~\bibnamefont{Weller}},
  \bibinfo{author}{\bibfnamefont{R.~J.} \bibnamefont{Bettles}},
  \bibinfo{author}{\bibfnamefont{P.}~\bibnamefont{Siddons}},
  \bibinfo{author}{\bibfnamefont{C.~S.} \bibnamefont{Adams}}, \bibnamefont{and}
  \bibinfo{author}{\bibfnamefont{I.~G.} \bibnamefont{Hughes}},
  \bibinfo{journal}{J. Phys. B: At. Mol. Opt. Phys.}
  \textbf{\bibinfo{volume}{44}}, \bibinfo{pages}{195006}
  (\bibinfo{year}{2011}).

\bibitem[{\citenamefont{Gibbs and Hull}(1967)}]{gibbs1967}
\bibinfo{author}{\bibfnamefont{H.~M.} \bibnamefont{Gibbs}} \bibnamefont{and}
  \bibinfo{author}{\bibfnamefont{R.~J.} \bibnamefont{Hull}},
  \bibinfo{journal}{Phys. Rev.} \textbf{\bibinfo{volume}{153}},
  \bibinfo{pages}{132} (\bibinfo{year}{1967}).

\bibitem[{\citenamefont{Agrawal}(2019)}]{agrawal}
\bibinfo{author}{\bibfnamefont{G.~P.} \bibnamefont{Agrawal}},
  \emph{\bibinfo{title}{Nonlinear Fiber Optics}} (\bibinfo{publisher}{Academic
  Press}, \bibinfo{year}{2019}).

\bibitem[{\citenamefont{Aladjidi et~al.}()\citenamefont{Aladjidi, Abuzarli,
  Brochier, Bienaimé, Picot, Bramati, and Glorieux}}]{aladjidi}
\bibinfo{author}{\bibfnamefont{T.}~\bibnamefont{Aladjidi}},
  \bibinfo{author}{\bibfnamefont{M.}~\bibnamefont{Abuzarli}},
  \bibinfo{author}{\bibfnamefont{G.}~\bibnamefont{Brochier}},
  \bibinfo{author}{\bibfnamefont{T.}~\bibnamefont{Bienaimé}},
  \bibinfo{author}{\bibfnamefont{T.}~\bibnamefont{Picot}},
  \bibinfo{author}{\bibfnamefont{A.}~\bibnamefont{Bramati}}, \bibnamefont{and}
  \bibinfo{author}{\bibfnamefont{Q.}~\bibnamefont{Glorieux}},
  \urlprefix\url{https://arxiv.org/abs/2202.05764}.

\bibitem[{\citenamefont{Siddons et~al.}(2008)\citenamefont{Siddons, Adams, Ge,
  and Hughes}}]{siddons2008}
\bibinfo{author}{\bibfnamefont{P.}~\bibnamefont{Siddons}},
  \bibinfo{author}{\bibfnamefont{C.~S.} \bibnamefont{Adams}},
  \bibinfo{author}{\bibfnamefont{C.}~\bibnamefont{Ge}}, \bibnamefont{and}
  \bibinfo{author}{\bibfnamefont{I.~G.} \bibnamefont{Hughes}},
  \bibinfo{journal}{J. Phys. B: At. Mol. Opt. Phys.}
  \textbf{\bibinfo{volume}{41}}, \bibinfo{pages}{155004}
  (\bibinfo{year}{2008}).

\end{thebibliography}

\clearpage

\onecolumngrid 



\section*{Supplementary material}

# Supporting Information
# Nonlinear refractive index of warm rubidium vapor

L. Kardum,[1] G. Premec,[1] N. Šantić,[1] and D. Aumiler[1]
[1] *Institute of Physics, Bijenička cesta 46, 10000 Zagreb, Croatia*
(Dated: June 12, 2026)

## I. TRANSIT TIME BROADENING

$\Gamma_t$ in Eqs. 2 and 3 in the main text accounts for the loss of atoms from the interaction zone with the laser beam due to their thermal motion in the vapor. This term includes transit time broadening effects into the model, where the broadening rate $\Gamma_t = 1/\tau_t$ is determined by the transit time $\tau_t$ of atoms through the laser beam. The atoms entering the interaction region with the beam are on the other hand in a statistical mixture of the two ground-state hyperfine levels, i.e. $\Gamma_{11} = \Gamma_t\, g_1/(g_1 + g_2)$ and $\Gamma_{22} = \Gamma_t\, g_2/(g_1 + g_2)$, where $g = 2F + 1$.

This model can be refined by considering the fraction of atoms that return inside the beam before loosing the internal state they have been prepared in before [1, 2]. After leaving the interaction area, atoms eventually interact with the cell walls and may return back to the interaction area in the same internal state. Paraffin coating of the cell walls may additionally conserve the atomic internal state after wall collisions [3]. Thus the influx rates $\Gamma_{11}$ and $\Gamma_{22}$ are not constant anymore but depend on the hyperfine level populations. $\Gamma_{22}$ is then given by:

$$\Gamma_{22} = (1-\beta)\frac{g_2}{g_1+g_2}\Gamma_t + \beta\left(\rho_{22} + \frac{\Gamma_{24}}{\Gamma_4}\rho_{44} + \frac{\Gamma_{25}}{\Gamma_5}\rho_{55} + \rho_{66}\right)\Gamma_t. \tag{1}$$

When $\beta = 0$, all the atoms have lost the memory of the state they have been prepared in before returning inside the beam and Eq. 1 reduces to $\Gamma_{22} = \Gamma_t\, g_2/(g_1 + g_2)$. If $\beta = 1$, the fraction of atoms entering the beam in $\rho_{11}$ and $\rho_{22}$ is given by the population diffusing out of the beam in those states plus a part of the population in the excited hyperfine levels, as the atoms leaving the beam in the excited state will decay into $\rho_{11}$ and $\rho_{22}$ before returning to the beam.

We note that $\beta = 0$ has been used in the calculations throughout the paper.

## II. HYPERFINE ABSORPTION SPECTRA OF RUBIDIUM VAPOR

We test our theoretical model in the linear regime by comparing it against the experimental results. Hyperfine absorption spectra of natural abundance rubidium vapor obtained near the room temperature presented in Ref. [4] are used. Transmission spectra are simulated using the calculated imaginary part of the complex refractive index using the relation:

$$\frac{I(L)}{I_0} = e^{-2k\, n_{Im}\, L}, \tag{2}$$

where $I_0$ and $I(L)$ are the laser intensity before and after passing through the vapor, respectively, and $L$ is the length of the vapor cell. The results are shown in Fig. 1, where the hyperfine absorption spectra at three different temperatures are compared to the simulations. Excellent quantitative agreement is achieved indicating that the theoretical model can be used to accurately model the linear complex refractive index of Rb vapor.

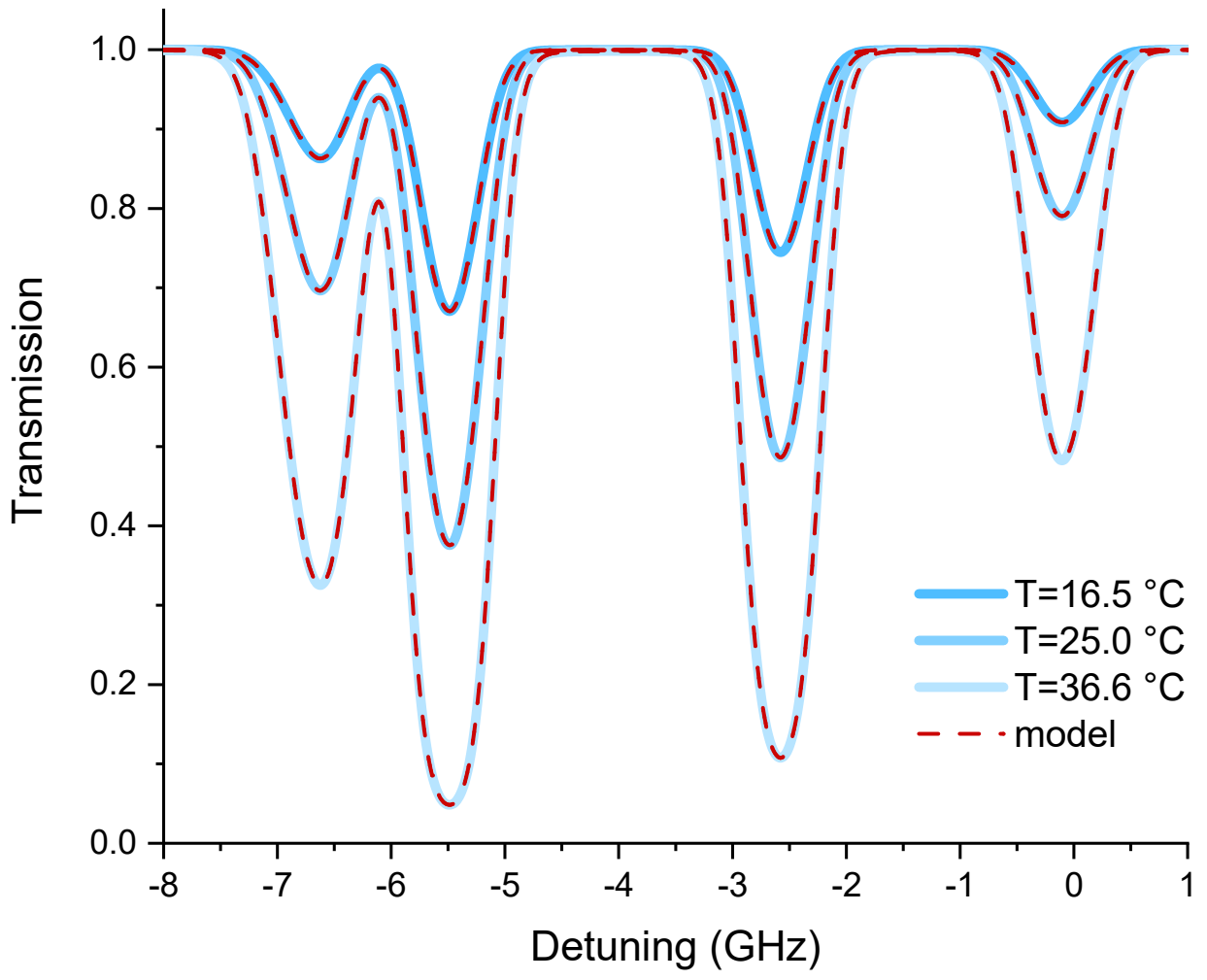


FIG. 1: Simulated hyperfine absorption spectra of rubidium D2 line at three vapor temperatures (dashed lines) compared to the experimental results from Ref. [4] (full lines). Two outer lines come from the $^{87}$Rb isotope, whereas the two inner lines come from the $^{85}$Rb isotope.

We note that an ab initio theoretical model with no free parameters is used, i.e. the only parameters used in the simulations are the experimental parameters describing the atomic vapor (cell temperature, which defines the atom concentration [5], and cell length), and the laser beam (laser power and beam cross section, which define the laser electric field and transit time broadening).

It takes about 5 seconds to calculate the transmission spectrum with 100 frequency data points using a standard desktop PC and using Python scripts available at `https://github.com/damiraumiler/Rb_D2_nonlinear_refractive_index`. This establishes our

model as a practical tool to model the complex refractive index of Rb vapor in real-life experimental environments, including the determination of Rb atom concentration from the measured hyperfine absorption spectra.

## III. NONLINEAR REFRACTIVE INDEX OF RUBIDIUM VAPOR FOR THE BLUE DETUNING

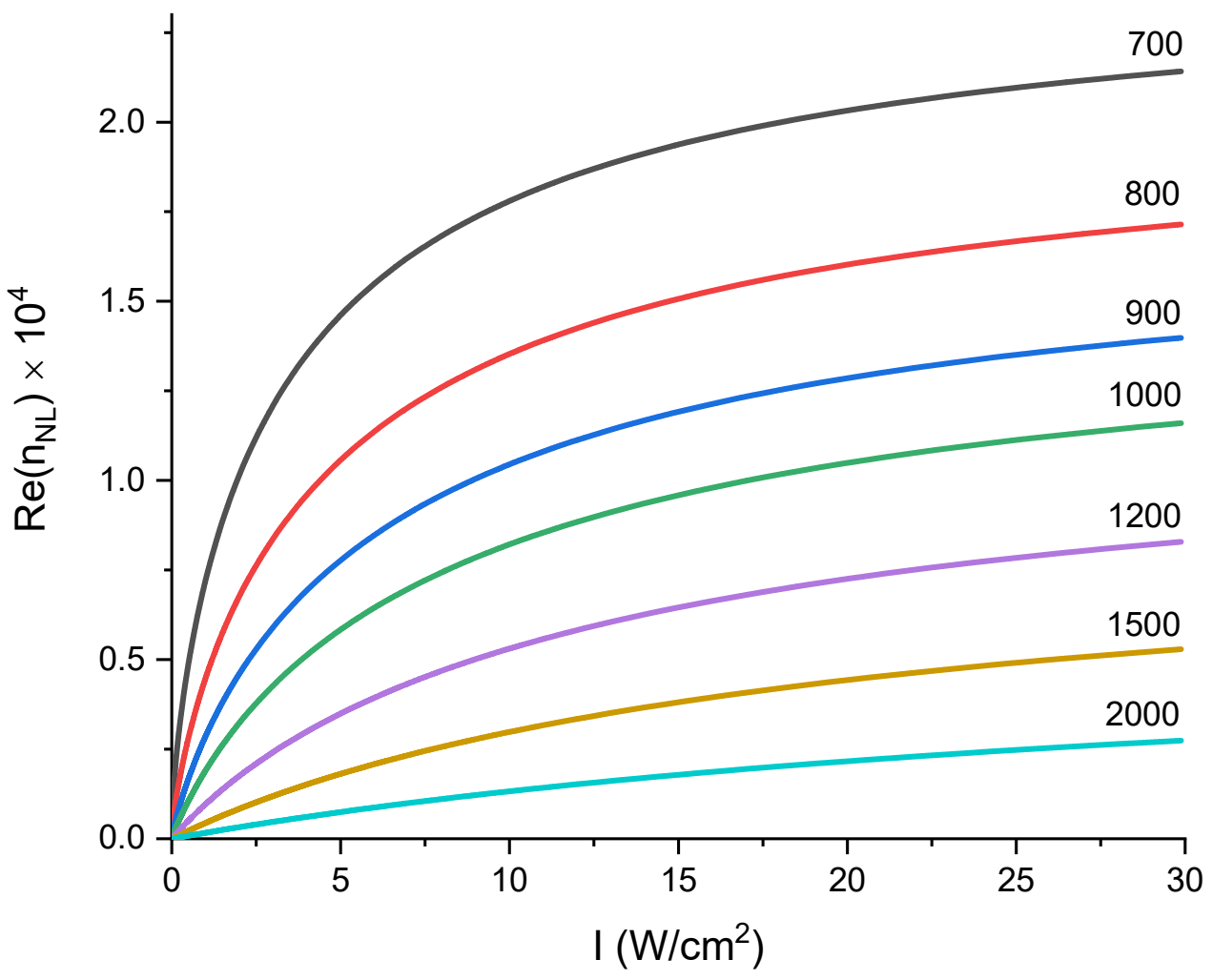


FIG. 2: Calculated real part of the nonlinear refractive index $n_{\mathrm{NL}}$ as a function of laser intensity for different laser detunings from the $^{87}$Rb $5S_{1/2}, F=1 \to 5P_{3/2}, F'=2$ transition (indicated in MHz). The parameters used in the calculations are the same as in Fig. 6 in main text.

We show in Fig. 2 the calculated nonlinear refractive index of $^{87}$Rb vapor for the same experimental parameters as in Fig. 6 in the main text and for the blue detuning. As the nonlinear refractive index shows clear saturation behavior with laser intensity, we can fit Eq. 1 in the main text to the calculated $n_{\mathrm{NL}}$ for different detunings. The resulting fitted parameters $n_2$ and $I_s$ are shown in Fig. 3 as a function of detuning.

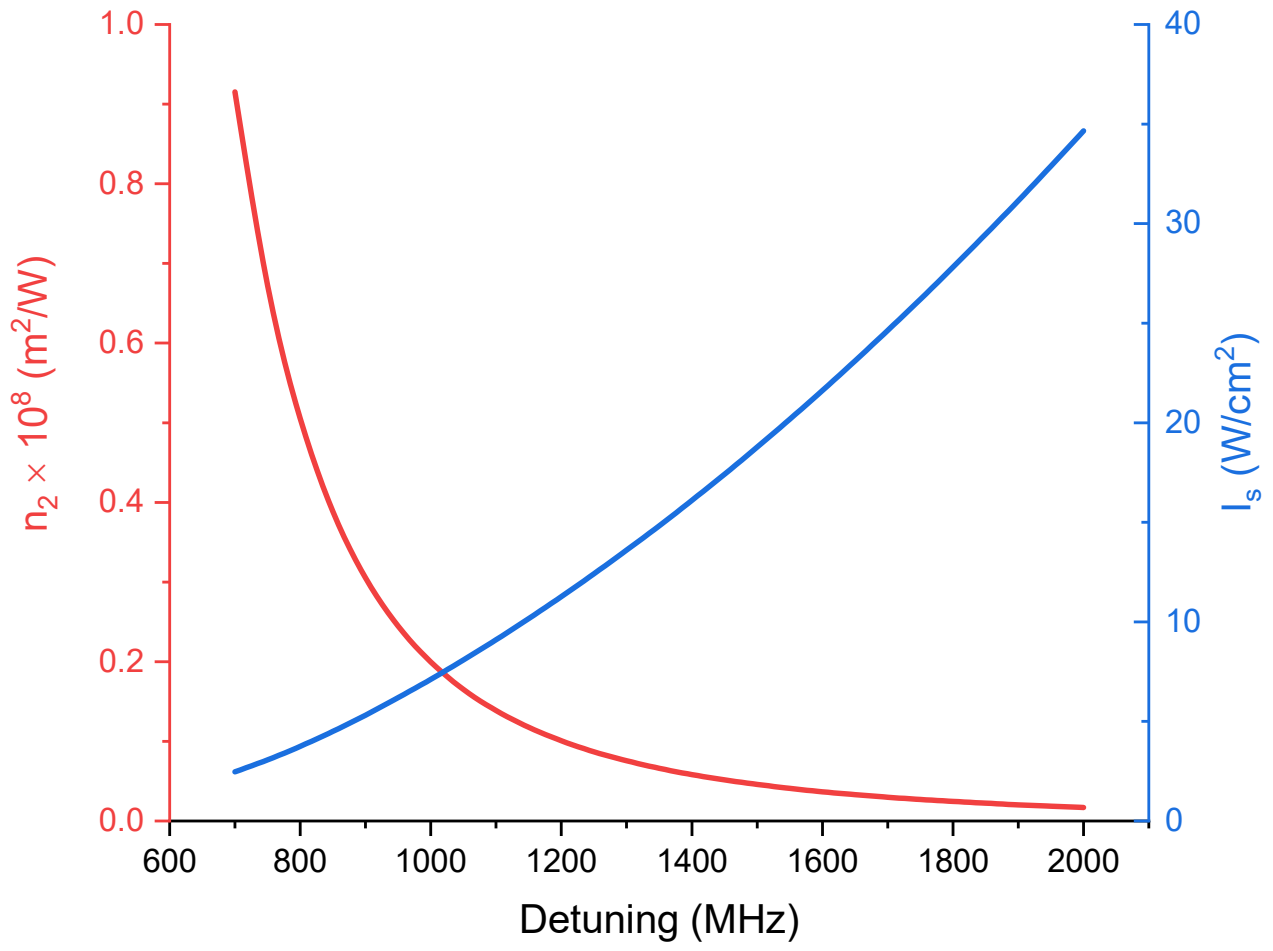


FIG. 3: Unsaturated Kerr coefficient $n_2$ (red) and nonlinear refractive index saturation intensity $I_s$ (blue) obtained by fitting the Eq. 1 in the main text to the results shown in Fig. 2 as a function of detuning from the $^{87}$Rb $5S_{1/2}, F=1 \to 5P_{3/2}, F'=2$ transition.

The results for the nonlinear refractive index of rubidium vapor for the blue detuning are both qualitatively and quantitatively similar to the results obtained for the red detuning presented in the main text.


[1] J. Ghosh, R. Ghosh, F. Goldfarb, J.-L. L. Gouët, and F. Bretenaker, Phys. Rev. A **80**, 023817 (2009).
[2] Q. Fontaine, Ph.D. thesis, Sorbonne Université (2020).
[3] M. Klein, M. Hohensee, D. Phillips, and R. Walsworth, Phys. Rev. A **83**, 013826 (2011).
[4] P. Siddons, C. S. Adams, C. Ge, and I. G. Hughes, J. Phys. B: At. Mol. Opt. Phys. **41**, 155004 (2008).
[5] A. N. Nesmeyanov, *Vapor Pressure of the Chemical Elements* (Elsevier, Amsterdam, 1963).



\end{document}